# Dynamic Time-Dependent Route Planning in Road Networks with User Preferences*

Moritz Baum[1], Julian Dibbelt[1], Thomas Pajor[2] and Dorothea Wagner[1]

[1]Karlsruhe Institute of Technology, Germany, `first.last@kit.edu`
[2]Sunnyvale, CA

December 30, 2015

## Abstract

There has been tremendous progress in algorithmic methods for computing driving directions on road networks. Most of that work focuses on time-independent route planning, where it is assumed that the cost on each arc is constant per query. In practice, the current traffic situation significantly influences the travel time on large parts of the road network, and it changes over the day. One can distinguish between traffic congestion that can be predicted using historical traffic data, and congestion due to unpredictable events, e. g., accidents. In this work, we study the *dynamic and time-dependent* route planning problem, which takes both prediction (based on historical data) and live traffic into account. To this end, we propose a practical algorithm that, while robust to user preferences, is able to integrate global changes of the time-dependent metric (e. g., due to traffic updates or user restrictions) faster than previous approaches, while allowing subsequent queries that enable interactive applications.

## 1 Introduction

The development of practical algorithms for route planning in road networks has been a showpiece application of successful algorithm engineering during the past decade. While Dijkstra's algorithm [16] solves the problem in almost linear time, it is too slow for responsive applications on real-world road networks of country or continental size. Thus, numerous *speedup techniques* have been proposed that employ a preprocessing phase to compute auxiliary data, which accelerates Dijkstra's algorithm by several orders of magnitude [1].

A successful approach [3, 7, 15, 20, 24] exploits that road networks have small separators [8, 21, 23, 32, 33]. Since this is a rarely changing topological feature, which is independent of the routing cost, preprocessing is split into two phases [7]: The first, metric-independent phase computes a nested hierarchy of separators and the overlay graph [34] induced by it. The second, metric-dependent phase (also called *customization*) computes distance-preserving weights for the predetermined overlay based on a given cost metric. Given the weighted overlay, queries can then skip large parts of the original graph during search. Customization turns out to be very fast: If the routing cost metric can be expressed as a single, scalar value, customization times below a second have been achieved on large continental networks [7, 15].

---

*Partially supported by EU grant 288094 (eCOMPASS) and EU grant 609026 (MOVESMART).

However, an important [13] aspect of route planning in realistic scenarios is the consideration of traffic patterns and incidents. For *dynamic, time-dependent* route planning, routing costs are no longer constant, scalar values per road segment. Rather, costs depend on the time of day at which a road is traversed, i.e., they vary as a function of time [5, 18]. Moreover, while these functions are typically precomputed from historic knowledge of traffic patterns [31], they have to be updated to respect current traffic incidents or short-term traffic predictions [14]. Also, the complexity of the functional description is known to strongly increase for long-distance routes [22], which poses a difficulty when adapting preprocessing techniques based on (long-distance) *shortcut edges* [12].

In this work, we investigate and experimentally evaluate the challenges that arise when extending a separator-based route planning approach to the dynamic time-dependent scenario.

**Related Work.** Dynamic shortest paths and time-dependency have been considered in previous works, often as separate problems. Dijkstra's algorithm [16] can be extended to compute time-dependent shortest paths [5, 18] if the time-dependent cost functions have certain, reasonable properties [30]. Two query variants are typically considered: (1) Given a source and target vertex and a fixed departure time at the source, compute the *earliest arrival time (EA)* at the target; (2) compute earliest arrival times for all departure times of a day (*profile search*). Time-dependent cost functions are typically represented as piecewise-linear functions, mapping departure to travel time. The complexity of such a function is expressed by its number of breakpoints. Foschini et al. [22] show that the optimal path between two vertices can change $n^{\Theta(\log n)}$ times, when varying departure time. In other words, for a given vertex pair, any representation of the functional dependency between departure time and travel time needs super-polynomial size. While these are worst-case bounds for arbitrary graph families, a very strong increase in functional complexity has also been observed by experimental evaluation of profile searches on real-world road networks [12].

Several known speed-up techniques have been adapted to handle time-dependency. Some works [10, 11, 29] use (scalar) lower bounds on the travel time functions to guide the graph search, circumventing any problems with functional complexity increase. TD-CALT [10] yields reasonable EA query times (at least for approximate solutions) and allows for fast dynamic traffic updates, but does not enable profile search on large networks. TD-SHARC [6] offers profile search on a country-scale network. Time-dependent Contraction Hierarchies (TCH) [2] enable both fast EA and profile searches even on continental networks. The technique (greedily) computes a vertex order, building a sequence of overlays by iteratively contracting "unimportant" vertices while inserting shortcuts between the remaining neighbors; The weights of shortcuts are obtained using profile searches. After preprocessing, piecewise-linear function approximation [25] can be used on the arc functions to save space, dropping optimality of subsequent query results. Quite cleverly, a multi-phase extension of the query algorithms (ATCH) restores exact results, despite approximated shortcuts [2]. Unfortunately, TCH and ATCH have not been shown to be robust against user preferences. In [28], time-dependent shortest path oracles are considered that provide approximate time-dependent distances in sublinear query time, after subquadratic preprocessing effort. The approach was also evaluated in experimental settings [27], but preprocessing effort and space consumption are rather high (even on a comparatively small, urban instance).

Other works consider unforeseen dynamic changes such as congestion after an accident. While this can trivially be solved by repeating the whole preprocessing phase, less costly approaches [11, 19] enable partial updates of preprocessed data. Customizable Route Planning (CRP) [7] goes even further, offloading most preprocessing effort to an initial metric-independent, separator-based phase. The preprocessed data is then *customized* to a given routing metric for the whole network within seconds or below (for scalar, non-functional metrics) in an additional preprocessing phase. Not only allows this fast dynamic changes, it also enables robust integration of user preferences and extended costs models such as turn-costs: Small graph separators in road networks guarantee efficient customization and query times for any scalar metric, since the induced overlay graphs are small. Customizable Contraction Hierarchies (CCH) [15] follows a similar approach.

However, for time-dependent metrics, small separators do not guarantee fast customization and query times: even having a low number of overlay arcs can be too expensive if the corresponding functional arc weights increase rapidly in complexity. To the best of our knowledge, non-scalar metrics for separator-based overlay approaches have so far only been investigated in the context of energy-optimal routes for electric vehicles (EVCRP) [4], where energy consumption depends on the battery state-of-charge, but functional complexity for long shortcuts grows very slowly.

On the other hand, it is important to note that the use of dynamic but scalar approaches (such as CRP or CCH) for handling live traffic information yields inaccurate results for medium and long-distance paths: Such methods will (wrongly) consider current traffic even at far away destinations—traffic that might well have dispersed once reaching the destination. For most-realistic results, a combination of dynamic and time-dependent (non-scalar, functional) route planning is necessary that accounts for current traffic, short-term predictions, and long-term historic knowledge about recurring traffic patterns.

**Contribution and Outline.** In this work, we carefully extend the Customizable Route Planning [7] approach to handle time-dependent functions. As such, we are the first to evaluate partition-based route planning on a challenging non-scalar metric. To this end, we integrate profile search into the customization phase and compute time-dependent overlays. We observe that, unlike EVCRP and TCH, a naïve implementation fails: Shortcuts on higher-level overlays become too expensive to be kept in memory (and too expensive to evaluate during queries). In order to reduce functional complexity, we propose to approximate overlay arcs after each level (in contrast to ATCH, which uses a post-processing step). In fact, even slight approximation suffices to make our approach practical, in accordance to theory [22]. The resulting algorithmic framework enables interactive queries with low average and maximum error in a very realistic scenario consisting of live traffic, short-term traffic predictions, and historic traffic patterns, while also supporting and being robust to user preferences such as lower maximum driving speeds or the avoidance of left-turns.

The rest of this paper is organized as follows. Section 2 introduces necessary notation and foundations of this work, while our approach is presented in Section 3. In Section 4, we provide an extensive experimental evaluation. We finish with concluding remarks in Section 5.

## 2 Preliminaries

A road network is modeled as a directed *graph* $G = (V, A)$ with $n = |V|$ *vertices* and $m = |A|$ *arcs*, where vertices $v \in V$ correspond to intersections and arcs $(u, v) \in A$ to road segments. An *s–t-path P* (in $G$) is a sequence $P_{s,t} = [v_1 = s, v_2, \ldots, v_k = t]$ of vertices such that $(v_i, v_{i+1}) \in A$. If $s$ and $t$ coincide, we call $P$ a *cycle*. Every arc $a$ has assigned a *travel-time function* $f_a \colon \Pi \to \mathbb{R}^+$ from a given function space $\mathbb{F}$, mapping departure time (within some period $\Pi = [0, \pi]$) to a positive, real-valued travel time. Given a departure time $\tau$ at $s$, the (time-dependent) travel time $\tau_{[s,\ldots,t]}$ of an $s$–$t$-path is obtained by consecutive function evaluation, i.e., $\tau_{[s,\ldots,v_i]} = f_{(v_{i-1},v_i)}(\tau_{[s,\ldots,v_{i-1}]})$. In what follows, we assume that $\mathbb{F}$ consists of *periodic piecewise linear functions* with period $\Pi$, given by *breakpoints*. For a function $f$, we denote by $|f|$ its number of breakpoints. Moreover, we define $f^{\max}$ as the maximum value of $f$, i.e., $f^{\max} = \max_{\tau \in \Pi} f(\tau)$. Analogously, $f^{\min}$ denotes the minimum value of $f$. A function $f$ is *constant* if $f \equiv c$ for some $c \in \Pi$. We presume that functions fulfill the *FIFO property*, that is, for arbitrary (positive) real values $\sigma \leq \tau \in \Pi$, the condition $\sigma + f(\sigma) \leq \tau + f(\tau)$ holds. In other words, waiting at a vertex never pays off. Note that the shortest path problem becomes $\mathcal{NP}$-hard if this condition is not satisfied for all arcs, unless waiting is allowed at vertices. On $\mathbb{F}$ we require binary *link* (composition) and *merge* operations. Given two functions $f, g \in \mathbb{F}$, the link operation is defined as $\mathrm{link}(f, g) := f + g \circ (\mathrm{id} + f)$, whereas we define merging $f$ and $g$ by $\mathrm{merge}(f, g) := \min(f, g)$. Given two piecewise linear functions $f, g$, the result of the link operation $\mathrm{link}(f, g)$ is piecewise linear again, with at most $|f| + |g|$ breakpoints (namely, at departure times of breakpoints of $f$ and backward projections of departure times of points of $g$). Similarly, the result of merging piecewise linear functions is piecewise linear, and the number of breakpoints is in $\mathcal{O}(|f| + |g|)$ (containing breakpoints of the two original functions and at most one intersection per linear segment). Both operations can be implemented by (coordinated) linear sweeps over the breakpoints of the corresponding functions.

Given a path $P = [v_1, \ldots, v_k]$, its *(travel-time) profile* is defined as the function $f_P \colon \Pi \to \mathbb{R}^+$ that maps departure time $\tau$ (at $v_1$) to travel time on $P$. Starting from $f_{[v_1, v_2]} = f_{(v_1, v_2)}$, we obtain the desired profile by consecutively applying the link operation, i.e., $f_{[v_1, \ldots, v_i]} = \mathrm{link}(f_{[v_1, \ldots, v_{i-1}]}, f_{(v_{i-1}, v_i)})$. Given a set of $s$–$t$-paths $\mathcal{P}$, the corresponding $s$–$t$-profile is defined by $f_\mathcal{P}(\tau) = \min_{P \in \mathcal{P}} f_P(\tau)$ for $\tau \in \Pi$, i.e., the *minimum profile* over all paths in $\mathcal{P}$. In other words, the profile $s$–$t$-profile maps departure time to minimum travel time for the given paths. Clearly, it can be obtained by (iteratively) merging the respective paths.

A *partition* of the vertices $V$ is a set $\mathcal{C} = \{C_1, \ldots, C_k\}$ of disjoint sets such that $\bigcup_{i=1}^k = V$. More generally, a *nested multi-level partition* consists of sets $\{\mathcal{C}^1, \ldots, \mathcal{C}^L\}$ such that $\mathcal{C}^\ell$ is a partition of $V$ for all $\ell \in \{1, \ldots, L\}$, and additionally for each cell $C_i$ in partition $\mathcal{C}^\ell$ of level $\ell < L$, there is a partition $\mathcal{C}^{\ell+1}$ of level $\ell + 1$ containing a cell $C_j$ such that $C_i \subseteq C_j$. We call $C_j$ the *supercell* of $C_i$. For the sake of consistency, we define $\mathcal{C}^0 = \{\{v\} \mid v \in V\}$ and $\mathcal{C}^{L+1} = \{V\}$. An arc $(u, v) \in A$ for which $u$ and $v$ are in different cells of $\mathcal{C}^\ell$ is called a *boundary arc* on level $\ell$. The vertices $u$ and $v$ are called *boundary vertices* (of level $\ell$).

**Query Variants and Algorithms.** Given a departure time $\tau$ and vertices $s$ and $t$, an *earliest-arrival (EA)* query asks for the minimum travel time (from $s$ to $t$) when departing at time $\tau$. Similarly, a *latest-departure (LD)* query asks (given vertices $s$, $t$, and an *arrival* time $\tau$) for the minimum travel time of an $s$–$t$-path reaching $t$ at time $\tau$. A profile query for given source $s$ and target $t$ asks for the minimum travel time at every possible departure time $\tau$, i.e., a profile $f_{s,t}$ from $s$ to $t$ (over all $s$–$t$-paths in $G$). EA queries can be handled by a time-dependent variant of Dijkstra's algorithm [18]. It maintains (scalar) *arrival time labels* $d(\cdot)$ for each vertex, initially set to $\tau$ for the source $s$ ($\infty$ for all other vertices). In each step, a vertex $u$ with minimum $d(u)$ is extracted from a priority queue (initialized with $s$). Then, the algorithm *relaxes* all outgoing arcs $(u, v)$: if $d(u) + f_{(u,v)}(d(u))$ improves $d(v)$, it updates $d(v)$ accordingly and adds (or updates) $v$ in the priority queue. Analogously, we can answer LD queries, by running the algorithm from $t$ and relaxing incoming instead of outgoing arcs, while maintaining departure time labels. We refer to this algorithm as *TD-Dijkstra*.

Profile queries can be solved by a *label-correcting* approach based on link and merge operations [12]. It generalizes Dijkstra's algorithm, storing (tentative) $s$–$v$ profiles $f_v$ at each vertex $v \in V$. Initially, we set $f_s \equiv 0$, and $f_v \equiv \infty$ for all other vertices. The algorithm continues along the lines of TD-Dijkstra, with a queue operating on scalar values $f_v^{\min}$ as key of a vertex $v$ with profile $f_v$. For extracted vertices $u$, arc relaxations propagate profiles rather than travel times. Hence, for each arc $(u, v)$, we compute $g := \mathrm{link}(f_u, f_{(u,v)})$ and set $f_v := \mathrm{merge}(f_v, g)$. If the profile $f_v$ changed, we update its key in the queue accordingly. We call this algorithm *Profile-Dijkstra*. As shown by Foschini et al. [22], the number of breakpoints of the profile of an $s$–$v$-paths can be superpolynomial, and hence, so is the space consumption *per vertex label* and the running time of Profile-Dijkstra in the worst case. Accordingly, Profile-Dijkstra is not feasible for large-scale instances, even in practice [12].

**Customizable Route Planning (CRP).** To allow fast integration of user-dependent routing preferences, CRP [7] introduces a three-phase workflow. The (metric-independent) preprocessing step of CRP consists of the computation of a multi-level partition of the input graph, for a given number of levels $L$. Several graph partition algorithms tailored to road networks exist, that provide partitions with balanced cell sizes and small cuts [8,23,32,33]. The multi-level partition is computed in an offline step and is therefore allowed to be (relatively) costly. For each level $\ell \in \{1, \ldots, L\}$, the resulting partition $\mathcal{C}^\ell$ induces an *overlay graph* $H^\ell$, containing all boundary vertices and boundary arcs in $\mathcal{C}^\ell$ and *shortcut* arcs between boundary vertices of each cell $C_i^\ell \in \mathcal{C}^\ell$. We define $\mathcal{C}^0 = \{\{v\} \mid v \in V\}$ and $H^0 := G$ for consistency. The topology (and the memory layout) of the overlay graphs (of each level) is constructed during preprocessing. This includes the allocation of data structures to store the metric-dependent data.

In the metric customization phase, costs of all shortcuts (added to the overlay graphs during preprocessing) are computed. This is achieved by running, for each boundary vertex of a given cell, Dijkstra's algorithm on the respective cell-induced subgraph. On higher levels, previously computed overlays are used for fast computation of shortcuts.

Finally, Multilevel Dijkstra [9, 24, 26] is used during the online phase to speedup queries. Given a source vertex $s$ and a target vertex $t$, the search graph consists of the overlay graph induced by the top-level partition $\mathcal{C}^L$, all overlays of cells of lower levels containing $s$ or $t$, and the level-0 cells (in the input graph $G$) that contain $s$ or $t$. Then, the query consists of a bidirectional variant of Dijkstra's algorithm, executed on this search graph.

## 3 Our Approach

In this section, we present *Time-Dependent CRP (TDCRP)*, a speedup technique for time-dependent queries that still allows fast integration of dynamic or user-dependent metric changes. In particular, TDCRP is able to incorporate long-term traffic patterns, live traffic, as well as short-term traffic predictions, while supporting preferences of users or vehicle restrictions (e. g., avoiding highways, maximum preferred driving speed, avoiding left-turns, vehicle height restrictions). Instead of scalar arc weights, we use functions of departure time at arcs, taking into account historic knowledge of traffic patterns.Additionally, we allow current and/or predicted traffic updates with a limited departure time horizon (i. e., accounting for the fact that underlying traffic situations resolve over time). This conceptual change has important consequences: While for (plain) CRP the topology of the overlays is fixed after preprocessing (enabling several micro-optimizations with significant impact on customization and query [7]), in our case the functional complexity is metric-dependent (influenced by, e. g., user preferences) and has to be handled dynamically during customization. This renders adaptation to dynamic time-dependent scenarios highly nontrivial, as we require new data structures and algorithmic changes during customization. Below, we describe our extensions to handle time-dependent scenarios in TDCRP. We incorporate *profile queries* into the customization phase to obtain *time-dependent* shortcuts. Moreover, we adapt the query phase to efficiently compute time-dependent shortest routes.

Regarding (metric-independent) preprocessing, on the other hand, the computation of a (multilevel) partition remains unchanged (compared to plain CRP). Building the overlay, we use the clique matrix representation, storing cliques of boundary vertices in matrices of contiguous memory [7]. However, instead of (scalar) arc weights, matrix entries now represent *pointers* to functions (whose complexity is not known until customization). To improve locality, all functions are stored in a single array, such that profiles corresponding to outgoing arcs of a certain boundary vertex are in contiguous memory. This dynamic data structure rules out several optimizations for (plain) CRP, such as microcode instructions, that require preallocated ranges of memory for the metric [7].

### 3.1 Customization

In the metric customization phase, we run profile searches to obtain costs of all time-dependent shortcuts. In particular, we require, for each boundary vertex $u$ (in some cell $C^\ell$ at level $\ell \geq 1$), the time-dependent distances for all $\tau \in \Pi$ to all boundary vertices $v \in C^\ell$. To this end, we run a profile query on the overlay $H^{\ell-1}$. By design, this query is restricted to *subcells* of $C^\ell$, i.e., cells $C^{\ell-1}$ on level $\ell - 1$ for which $C^{\ell-1} \subseteq C^\ell$ holds. This yields profiles for all outgoing (shortcut) arcs $(u, v)$ in $C^\ell$ from $u$. Unfortunately, profile queries are expensive in terms of both running time and space consumption. Below, we describe improvements to remedy these effects, mostly by tuning the profile searches.

**Improvements.** The main bottleneck of the profile search is performing link and merge operations, which require linear time in the function size (cf. Section 2). To avoid unnecessary operations, we explicitly compute and store the minimum $f^{\min}$ and the maximum $f^{\max}$ of each profile $f$ in its corresponding label (and in shortcuts of overlays). These values can be used for early pruning (while avoiding costly link and merge operations): before relaxing an arc $(u, v)$, we check whether $f_u^{\min} + f_{(u,v)}^{\min} > f_v^{\max}$, i.e., the minimum of the linked profile exceeds the maximum of the label at vertex $v$. If this is the case, the arc $(u, v)$ does not need to be relaxed. Otherwise, the functions are linked. We distinguish four cases, depending on whether the first and second function are constant, respectively. If both functions are constant, linking becomes trivial (summing up two integers). If one of them is constant, simple shift operations suffice (note that we need to distinguish two cases, depending on which of the two functions is constant). Only if none of the functions is constant, we apply the actual link operation with running time linear in $|f_u| + |f_{(u,v)}|$.

After linking $f_{(u,v)}$ to $f_u$, we obtain a tentative label $\tilde{f}_v$ together with its minimum $\tilde{f}_v^{\min}$ and its maximum $\tilde{f}_v^{\max}$. We perform additional checks before merging the profiles $f_v$ and $\tilde{f}_v$ in order to avoid unnecessary expensive merge operations. First, we perform bound checks: if $\tilde{f}_v^{\min} > f_v^{\max}$, the function $f_v$ remains unchanged, and no merge is necessary (note that this situation may occur even though we checked the bounds before linking). Conversely, if $f_v^{\max} < \tilde{f}_v^{\min}$, we can simply replace $f_v$ by $\tilde{f}_v$. Second, if both checks fail and the functions $f_v$ and $\tilde{f}_v$ are both nonconstant, we perform an additional check to determine whether merging is necessary (still, one function might dominate the other). This check consists of a simulated merge operation (thus, a coordinated linear-time sweep over the breakpoints of each function), which omits (computationally involving and numerically unstable) line segment intersections.

Additionally, we can maintain a special value per vertex, indicating its *unique parent* (i.e., the distinct vertex contributing to the profile at some vertex) if it exists. (Recall that for different points in time, the optimal path to a vertex, and thus its parent, may vary.) We use this value to perform *hopping reduction* [17]. If at some point, before relaxing the arc $(v, u)$, $u$ is the unique parent of $v$ (i.e., the label at $v$ has not been updated since this arc was relaxed), we can safely omit linking and merging (because a shortest path cannot contain a cycle). Alternatively, we can use more powerful *clique flags*: For each vertex of the overlay graph, we add a flag to its label that is true if *all* its parents belong to the same cell (on the current level). This flag is set to true whenever the corresponding vertex label $f_v$ is set to the tentative function $\tilde{f}_v$ after relaxing a clique arc $(u, v)$ (i.e., the label is set for the first time or the former label $f_v$ is dominated by the tentative function $\tilde{f}_v$). It is set to false if the vertex label is (partially) improved after relaxing a boundary arc. For flagged vertices, we do not relax outgoing clique arcs, as this cannot possibly improve labels within the same cell (due to the triangle inequality and the fact that we use full cliques).

**Parallelization.** Since cells on a given level are processed independently, CRP can be parallelized in a straightforward manner, assigning cells to different threads [7]. In our scenario, however, the workload may differ significantly between cells, as it strongly correlates with the number of time-dependent arcs in the search graph. (Conversely, cell sizes are the bounding factor in plain CRP, which are balanced by the partition.) As an extreme example, a cell that contains no time-dependent arcs at all reduces profile searches to simple Dijkstra runs. In realistic data sets, the distribution of time-dependent arcs is clearly not uniform: it depends on the road type (e.g., highways vs. side roads) and the area (rural vs. urban). To balance the load per thread, we no longer parallelize per cell, but per boundary vertex.

In contrast to plain CRP, shortcut profiles are written to dynamic containers, since the number of breakpoints is not known in advance. Therefore, we have to prohibit parallel (writing) access to the respective data structure. One way to solve this is to make use of locks. However, this can be expensive if many threads try to write profiles at the same time. Alternatively, we can use thread-local profile containers (i.e., each thread uses its own container to store profiles). After customization of each level, we synchronize data by copying profiles to the global container sequentially. For better locality during queries, we maintain the relative order of profiles (wrt. the matrix layout, so profiles of adjacent vertices are likely to be contiguous in memory). Since the relative order of profiles in thread-local containers is ensured by running queries accordingly, running merge sort on these containers is sufficient to achieve the global order.

**Approximations.** On higher levels of the partition, shortcuts represent larger parts of the graph and contain more breakpoints. Accordingly, shortcuts on higher levels are more expensive (wrt. memory consumption). This makes profile searches fail on large graphs due to insufficient memory, even on modern hardware. Moreover, running time is strongly correlated to the complexity of profiles. In order to save space and time, we can *simplify* functions during customization. To this end, we use the algorithm by Imai and Iri [25] that, given a maximum (relative or absolute) error $\varepsilon$, computes an approximation of a given piecewise linear function with a provably minimum number of breakpoints. In [2], this technique is applied after preprocessing to reduce space consumption. Instead, we use this algorithm to simplify profiles after computing all shortcuts of a certain level. Therefore, searches on higher levels use approximated functions from lower levels, leading to (slightly) less accurate profiles but faster customization. The parameter $\varepsilon$ is a tuning parameter: larger values allow faster customization, but decrease quality. Also, approximation is not necessarily applied on all levels, but can be restricted to the higher ones. Note that, when using clique flags on approximated shortcuts, the triangle inequality may no longer hold (due to the fact that bounds were simplified after computing shortcuts). Therefore, clique flags yield faster (profile) queries and (slightly) decreased quality (additional arc relaxations improve shortcut bounds).

## 3.2 Live Traffic and Short-Term Traffic Predictions

Updates due to, e.g., live traffic, require that we rerun the customization phase. Clearly, we only have to perform customization only for *affected* cells, i.e., cells that contain at least one arc for which an update is made. However, we can do better if we exploit the fact that live traffic updates are not periodic. Also, short-term updates only affect a limited time horizon (rather than the complete period). Thus, we do not have to propagate updates to boundary vertices that cannot reach the affected arc before the end of the time horizon.

We assume that short-term updates are given as *partial functions* $f\colon [\pi', \pi''] \to \mathbb{R}^+$, where $\pi' \in \Pi$ and $\pi'' \in \Pi$ are the *beginning* and *end* of the time horizon, respectively. Let $a_1 = (u_1, v_1), \ldots, a_k = (u_k, v_k)$ denote the *updated* arcs inside some cell $C^\ell$ at level $\ell$, and let $f_1, \ldots, f_k$ be the corresponding partial functions representing time horizons. Moreover, let $\tau$ be the current point in time. To update $C^\ell$ we run, on its induced subgraph, a *multi-target* latest-departure query from the tails of all updated arcs. In other words, we initially insert the vertices $u_1, \ldots, u_k$ into the queue. The departure time label of $u_i$ (for $i \in \{1, \ldots, k\}$) is set to $\pi_i''$ (i.e., the end of the time horizon $[\pi_i', \pi_i'']$ of the partial function $f_i$). Consequently, the LD query computes, for each vertex of the cell $C^\ell$, the latest possible departure time such that some affected arc is reached on a shortest path (before the end of its corresponding time horizon). Whenever the search reaches a boundary vertex of the cell, it is marked as *affected* by the update. We stop the search as soon as the departure time label of the current vertex is below $\tau$ (recall that the LD query visits vertices in decreasing order of departure time). Thereby, we ensure that only such boundary vertices are marked from which an updated arc can be reached in time.

Next, we run the actual profile searches for $C^\ell$ (similar to regular customization), but only from affected vertices. For all shortcut profiles obtained during these searches, we test whether they improve the corresponding stored shortcut. If this is the case, we add the interval of the profile for which a change occurs to the set of time horizons for the next level. If shortcuts are approximations, we test whether the change is *significant* (e.g., if the maximum difference between the profiles exceeds a certain bound). We continue the update process on the next level accordingly.

## 3.3 Queries

The query algorithm makes use of shortcuts computed during customization to reduce the search space. Similar to plain CRP, the search graph for a given source $s$ and target $t$ consists of the overlay graph induced by the top-level partition $\mathcal{C}^L$ and subgraphs induced by cells containing $s$ or $t$. This search graph does not have to be constructed explicitly, but can be obtained implicitly by obtaining, at each vertex $v$, the highest *uncommon* level of $v$ and both $s$ and $t$ in the multi-level partition: We compute the highest levels $\ell_{s,v}$ and $\ell_{v,t}$ of the partition such that $v$ and $s$ or $t$, respectively, are in distinct cells of the partition (0 if $v$ is in the same level-1 cell as $s$ or $t$). Then, we relax only outgoing edges of $v$ at level $\min\{\ell_{s,v}, \ell_{v,t}\}$ (recall that $H^0 = G$). Since the search graph is known at query time, we can simply run (unidirectional) TD-Dijkstra to answer earliest arrival queries.

We also make use of the minimum values $f_{(u,v)}^{\min}$ stored at arcs during EA queries: If, at some point during the query, $d(u) + f_{(u,v)}^{\min}$, does not yield an improvement of $d(v)$, we do not have to relax the arc $(u,v)$. Thereby, we avoid costly function evaluation. For approximated queries (i.e., if we use approximated shortcuts), we observe that clique flags improve running times particularly when activated on higher levels (where cliques are larger). On the other hand, we observe quality issues with activated clique flags mostly on the lower levels (since errors propagate to higher levels). Hence, we activate clique flags only on the top-$k$ levels of the partition (where $k$ is a tuning parameter).

## 3.4 Implementation Details

Implementation of the functional operations (evaluation, link, merge, and subroutines) requires considerable effort and is a great source for numerical imprecision. We tested several implementation variants and settled on using fixed-point arithmetics at milliseconds resolution. While specifics are beyond the scope of this paper, we note that we found it beneficial to use half-plane tests (which require no division) to determine whether an intersection between line segments occurs, before computing the actual intersection.

We reorder vertices of the graph, moving boundary vertices to the front, and ordering vertices of the same level by cell IDs [7]. This improves locality of subsequent memory accesses. In a naive implementation, each vertex requires its own vertex label. Since the search graph is known in advance, we can reduce the number of vertex labels to save space. Instead of explicitly extracting the search graph, we compute the ranges of IDs of vertices of the current cell for each level (note that there is at most one range per level, due to vertex reordering). Then, we can remap the ranges of each level (for source and target cell, respectively) to a smaller, global range of vertex indices. The size of the global range depends on the maximum cell size, which is known after preprocessing. The following (mixed) variant worked best in our experiments: We only remap bottom level inner-cell indices (the majority of vertices), while keeping a distinct vertex label for every boundary vertex in the graph. Thereby, we save a significant amount of space, improve locality, but keep vertex mapping overhead limited during queries.

To reset labels between queries, rather than using standard approaches (like timestamps), we again exploit the vertex reordering: We store, for every cell, its vertex ranges, and reset only the (at most two) cells per level touched during a query, along with all top level vertices. With labels being on adjacent ranges, this can be done efficiently in practice.

Finally, we can save some space by storing cell IDs only at boundary vertices. Before running the actual query algorithm, the source and target cell can be retrieved by running an additional DFS in their respective (bottom-level) cells at negligible overhead.

## 4 Experiments

We implemented all algorithms in C++ using g++ 4.8 (flag -O3) as compiler. Experiments were conducted on a dual 8-core Intel Xeon E5-2670 clocked at 2.6 GHz, with 64 GiB of DDR3-1600 RAM, 20 MiB of L3 and 256 KiB of L2 cache. We ran customization in parallel (using all 16 threads) and queries sequentially.

**Input Data and Methodology.** Our test instances are based on the road network of Germany ($n = 4.7$ million, $m = 10.8$ million) and Western Europe ($n = 18$ million, $m = 42.2$ million), kindly provided by PTV AG, and the road network of Berlin/Brandenburg ($n = 443\,k$, $m = 988\,k$), kindly provided by TomTom. All inputs contain time-dependent data. For the Germany and Berlin instance, data stems from historical traffic patterns. For both instances, we extracted the 24 hour profile of a Tuesday. For Western Europe, travel time functions were generated synthetically [29]. See Table 1 for details on inputs. For partitioning, we used PUNCH [8], which is explicitly developed for road networks and aims at minimizing the number of boundary arcs. For Germany, we use a 5-level partition, with maximum cell sizes of $2^{[4:8:12:15:18]}$. For Europe, we use a 6-level partition, with maximum cell sizes $2^{[4:8:11:14:17:20]}$. Finally, we use a 5-level partition for Berlin, with cell size restricted to $2^{[4:8:11:14:17]}$. Compared to plain CRP, we use partitions with more levels, to allow fine-grained approximation. Computing the partition took 20 seconds for Berlin, 5 minutes for Germany, and 23 minutes for Europe. Given that road topology changes rarely, this is sufficiently fast in practice.

**Table 1.** Network properties. We report the number of vertices and arcs of the routing graph, and as a measure of time-dependency, the total amount of break points in the whole network as well as the average time-dependent arc complexity.

| Network | # Vertices | # Arcs | Time-dep. arcs | # Break points Total | Avg. |
|---|---:|---:|---:|---:|---:|
| Berlin  |    443 191 |    988 493 |   271 688 (27.5%) |  3 464 241 | 12.8 |
| Germany |  4 692 091 | 10 805 429 |   777 984 (7.2%)  | 12 714 370 | 16.3 |
| Europe  | 18 010 173 | 42 188 664 | 2 609 651 (6.2%)  | 29 362 693 | 11.3 |

**Table 2.** Custom performance on Europe instance for varying approximation errors ($\varepsilon$). We report, per level, the number of breakpoints (bps, in millions) in the resulting overlay, the percentage of clique arcs that are time-dependent (td.clq.arcs), the average complexity of time-dependent arcs (td.arc.cplx), as well as the customization time for that level. Without approximation, level 5 and 6 cannot be computed as they do not fit into main memory.

| $\varepsilon$ | | Lvl1 | Lvl2 | Lvl3 | Lvl4 | Lvl5 | Lvl6 | Total |
|---|---|---|---|---|---|---|---|---|
| — | bps [$10^6$] | 99.1 | 398.4 | 816.4 | 1 363.4 | — | — | 2 677.4 |
| | td.clq.arcs [%] | 17.0 | 52.6 | 76.0 | 84.2 | — | — | — |
| | td.arc.cplx | 21.0 | 68.9 | 189.0 | 509.3 | — | — | — |
| | time [s] | 11.4 | 52.0 | 152.9 | 206.2 | — | — | 375.7 |
| 0.01% | bps [$10^6$] | 75.6 | 182.7 | 244.6 | 240.8 | 149.3 | 59.2 | 952.2 |
| | td.clq.arcs [%] | 16.9 | 52.6 | 75.9 | 84.1 | 85.1 | 82.4 | — |
| | td.arc.cplx | 16.0 | 31.6 | 56.6 | 89.9 | 108.6 | 108.0 | — |
| | time [s] | 4.5 | 18.2 | 33.1 | 83.4 | 152.2 | 153.3 | 444.7 |
| 0.1% | bps [$10^6$] | 60.7 | 107.5 | 111.5 | 87.9 | 47.9 | 17.5 | 432.9 |
| | td.clq.arcs [%] | 17.0 | 52.7 | 76.0 | 84.2 | 85.2 | 82.5 | — |
| | td.arc.cplx | 12.9 | 18.6 | 25.8 | 32.8 | 34.8 | 32.1 | — |
| | time [s] | 4.2 | 16.1 | 21.6 | 41.2 | 63.2 | 55.7 | 202.0 |
| 1.0% | bps [$10^6$] | 45.6 | 58.0 | 45.6 | 29.2 | 14.7 | 5.3 | 198.5 |
| | td.clq.arcs [%] | 17.0 | 52.7 | 76.0 | 84.2 | 85.2 | 82.5 | — |
| | td.arc.cplx | 9.7 | 10.0 | 10.6 | 10.9 | 10.7 | 9.8 | — |
| | time [s] | 4.1 | 14.1 | 14.9 | 22.85 | 30.0 | 24.3 | 110.2 |

**Evaluating Customization.** Table 2 shows details on customization for different approximation parameters $\varepsilon$ on the Europe instance. We report, for several choices of $\varepsilon$ (and for each level of the partition), the total number of breakpoints, the ratio of time-dependent (i.e., non-constant) shortcuts, the average number of breakpoints per time-dependent shortcuts, and the (parallelized) customization time. The first block shows figures for exact profile computation. Customization had to be aborted after the fourth level, because the 64 GiB of main memory were not sufficient to store all vertex (profile) labels. For remaining levels, we clearly see the strong increase of total breakpoints per level. Also, the (relative) amount of time-dependent arcs rises with each level (since shortcuts become longer). Finally, customization time clearly correlates with profile complexity, from 10 seconds on the lowest level, to more then three minutes on the fourth. When approximating, we clearly see that customization becomes faster for larger values of $\varepsilon$. Recall that higher levels work on approximated shortcuts of previous levels, so $\varepsilon$ is not an approximation bound for all shortcuts. Still, even a very small value (0.01 %) yields a massive drop of profile complexity (more than a factor 5 at level 4), and immediately allows complete customization. For reasonably small values ($\varepsilon = 0.1\,\%, \varepsilon = 1.0\,\%$), we see that customization becomes much faster (less than two minutes for $\varepsilon = 1.0\,\%$) and in particular fast enough for real-time updates. Even for larger values of $\varepsilon$, the higher levels are far more expensive (due to the large amount of time-dependent arcs).

**Table 3.** Single-core query performance on Europe instance as a trade-off of customization effort and approximation. Since we employ approximation per level, the resulting query errors can be higher than the input parameter.

| Customization | | Query | | | | | |
|---|---|---|---|---|---|---|---|
| Time[s] | $\varepsilon$ | #Vertices | #Arcs | #Bps | Time [ms] | Avg. err. [%] | Max. err. [%] |
| 444.7 | 0.01% | 3 499 | 12 231 | 347 559 | 6.28 | <0.01 | 0.08 |
| 202.0 | 0.1% | 3 499 | 12 248 | 78 404 | 4.02 | 0.04 | 0.81 |
| 110.2 | 1.0% | 3 499 | 12 259 | 54 784 | 3.65 | 0.54 | 3.21 |

**Evaluating Queries.** Table 3 details query performance for different values of the approximation parameter $\varepsilon$ on the Europe. For each, we report the number of settled vertices, arc relaxations, and total number of evaluated breakpoints during the query. Furthermore, we report timings, average, and maximum error for 100 000 point-to-point queries. In each query, source and target vertex, as well as departure time were chosen uniformly at random. The data shows that (similar to customization), query times decrease with higher approximation ratio. Again, this is due to the smaller number of breakpoints in profiles. As expected, the average and maximum error increase with $\varepsilon$. However, we see that even for the largest value $\varepsilon = 1.0\,\%$, the maximum error is very low (about 3 %, which should be acceptable in most reasonable scenarios). Moreover, query times are quite practical for all values of $\varepsilon$, ranging from less than 4 ms to below 7 ms. In summary, our approach allows query times that are fast enough for interactive applications, if a reasonably low error is allowed. Given that the input functions are based on statistical input with inherent inaccuracy, this should be more than acceptable for most applications.

**Comparison with Related Work.** Finally, Table 4 provides an overview of our approach compared to the most relevant previous work on time-dependent routing. For each technique, we report performance of metric-dependent preprocessing and EA queries. We evaluated our approach on all benchmark instances, namely, Berlin, Germany, and Europe (each for the fastest variant, i.e., $\varepsilon = 1.0$). For every approach (and instance), the table shows (metric-dependent) preprocessing time (and the number of cores), as well as space consumption. For EA queries, we report vertex scans, query time, and (average and maximum) errors. We see that our approach competes with all related approaches, but has much lower customization time. Accounting for differences in hardware and implementation, TDCALT may possibly be preprocessed much faster (also, running preprocessing in parallel should be possible). This makes TDCALT the only candidate that might compete with TDCRP in terms of preprocessing effort and resulting query performance. However, without TDCALT allows no profile searches, making it a less versatile approach. Also, the support of advanced features, such as short-term traffic predictions, has not beed discussed for TDCALT. All in all, we see that TDCRP is a competitive approach that clearly broadens the state-of-the-art of time-dependent route planning, as it handles a wider range of practical requirements.

**Table 4.** Overview of related work. For Germany and Europe, we present figures (as-is) for TDCALT [10] (evaluated on an Opteron 2218 at 2.67 GHz), SHARC [6] (Opteron 2218, 2.67 GHz), TCH and ATCH [2] (2×4×Core-i7, 2.67 GHz). For Berlin, we report measures for FLAT [28] (machine not specified). Also, instances are slightly different for Berlin. For TCH and ATCH, preprocessing times distinguish vertex ordering and contraction.

| Algorithm | Instance | Metric-dep. Prepro. [h:m:s] (# Thr) | Space [B/n] | EA Queries # Vert. | Time [ms] | Err. [%] avg. | Err. [%] max. |
|---|---|---|---|---|---|---|---|
| TDCALT | Germany | 9:00 (1) | 50 | 3 190 | 5.36 | — | — |
| TDCALT-K1.15 | Germany | 9:00 (1) | 50 | 1 593 | 1.87 | 0.05 | 13.84 |
| eco L-SHARC | Germany | 1:18:00 (1) | 219 | 2 776 | 6.31 | — | — |
| heu SHARC | Germany | 3:26:00 (1) | 137 | 818 | 0.69 | n/a | 0.61 |
| TCH | Germany | 5:04 + 1:14 (8) | 995 | 520 | 0.75 | — | — |
| ATCH (1.0) | Germany | 5:04 + 1:14 (8) | 239 | 588 | 1.24 | — | — |
| inex. TCH (0.1) | Germany | 5:04 + 1:14 (8) | 286 | 642 | 0.70 | 0.02 | 0.10 |
| inex. TCH (1.0) | Germany | 5:04 + 1:14 (8) | 214 | 654 | 0.69 | 0.27 | 1.01 |
| inex. TCH (2.5) | Germany | 5:04 + 1:14 (8) | 172 | 668 | 0.72 | 0.79 | 2.44 |
| inex. TCH (10.0) | Germany | 5:04 + 1:14 (8) | 113 | 898 | 1.06 | 3.84 | 9.75 |
| TDCRP (1.0) | Germany | 0:08 (16) | 77 | 2 152 | 1.17 | 0.68 | 3.60 |
| TDCALT | Europe | 1:00:00 (1) | 61 | 60 961 | 121.4 | — | — |
| TDCALT-K1.05 | Europe | 1:00:00 (1) | 61 | 32 405 | 62.5 | 0.01 | 3.94 |
| TDCALT-K1.10 | Europe | 1:00:00 (1) | 61 | 12 777 | 21.9 | 0.09 | 7.88 |
| TDCALT-K1.15 | Europe | 1:00:00 (1) | 61 | 6 365 | 9.2 | 0.26 | 8.69 |
| eco L-SHARC | Europe | 6:49:00 (1) | 198 | 18 289 | 38.29 | — | — |
| heu SHARC | Europe | 22:12:00 (1) | 127 | 5 031 | 2.94 | n/a | 0.61 |
| TCH | Europe | 37:42 + 8:02 (8) | 599 | 1 021 | 2.11 | — | — |
| ATCH (1.0) | Europe | 37:42 + 8:02 (8) | 208 | 1 223 | 2.89 | — | — |
| inex. TCH (0.1) | Europe | 37:42 + 8:02 (8) | 239 | 1 722 | 2.70 | 0.02 | 0.15 |
| inex. TCH (1.0) | Europe | 37:42 + 8:02 (8) | 195 | 1 782 | 2.76 | 0.20 | 1.50 |
| inex. TCH (2.5) | Europe | 37:42 + 8:02 (8) | 175 | 1 875 | 2.94 | 0.48 | 3.37 |
| inex. TCH (10.0) | Europe | 37:42 + 8:02 (8) | 144 | 1 801 | 2.92 | 2.88 | 16.21 |
| TDCRP (1.0) | Europe | 1:50 (16) | 133 | 3 499 | 3.65 | 0.54 | 3.21 |
| FLAT $K_{2000}$ | Berlin* | 4:30:00 (6) | 58 000 | 1 522 | 0.55 | 0.24 | n/a |
| FLAT $SR_{2000}$ | Berlin* | 4:30:00 (6) | 58 000 | 120 | 0.07 | 0.73 | n/a |
| TDCRP (1.0) | Berlin | 0:01 (16) | 67 | 854 | 0.28 | 1.47 | 2.69 |

# 5 Conclusion

In this work, we introduced TDCRP, an extension of CRP to time-dependent query scenario. We showed how integrating profile searches into the customization phase of CRP allows us to compute time-dependent overlays. Thereby, we enable integration of dynamic or user-dependent metric changes. Additionally, TDCRP also efficiently incorporates live traffic and short-term traffic predictions. Our experimental analysis on a continental road network shows that for higher-level overlays (i. e., where overlay vertices are far apart), the increase in functional complexity is too high to be kept in memory (both in contrast to the known techniques EVCRP and TCH). To reduce memory consumption, we approximate the overlay arcs at each level, accelerating the customization and query times. As a result, we obtain an approach that enables fast near-optimal queries, while being able to quickly integrate user preferences, live traffic updates and traffic predictions.

There are several aspects of future work. First, we would like to further investigate TDCALT, which appears quite competitive in scenarios where profile searches are not a requirement. For example, TDCALT should be easy to parallelize, so a re-evaluation might be fruitful (e. g., exploiting insights from [19]). Additionally, we are interested in alternative customization approaches (avoiding expensive profile searches). This could be achieved, e. g., by making use of kinetic data structures [22], or balanced contraction within cells. As demonstrated for ATCH, one could also aim at exact queries based on approximated shortcuts, by employing additional query phases.

**Acknowledgements.**   We thank Gernot Veit Batz, Daniel Delling, Felix König, Spyros Kontogiannis, and Ben Strasser for interesting conversations.